%% file: main.tex
\documentclass[a4paper,UKenglish,cleveref, pdfa, autoref, thm-restate]{lipics-v2021}

\usepackage{pifont}
\usepackage[utf8]{inputenc}
\usepackage{xcolor}
\usepackage{lstcoq}
\hideLIPIcs
\nolinenumbers

\newcommand{\HTMLBase}{https://FormalML.github.io/ITP22/documentation/html}
\newcommand{\coqHTMLBase}{\HTMLBase}
\newcommand{\coqBaseModule}{FormalML.}

\newcommand{\fleur}{\ding{95}}
\newcommand{\coqtop}{\text{\href{https://github.com/IBM/FormalML}{\fleur}}}
\newcommand{\coqdef}[2]{\text{\href{\coqHTMLBase/\coqBaseModule#1.html\##2}{\fleur}}}
\pdfoutput=1 

\title{Formalization of a Stochastic Approximation Theorem}

\author{Koundinya {Vajjha}}{University
  of Pittsburgh, United States \and
  \url{https://kodyvajjha.github.io/}}{kov5@pitt.edu}{https://orcid.org/0000-0003-3799-6326}{Vajjha
  acknowledges support from the Alfred P. Sloan Foundation under grant number
  G-2018-10067 and the Andrew W. Mellon Foundation.}

\author{Barry Trager}{IBM Research, United States}{bmt@us.ibm.com}{}{}

\author{Avraham Shinnar}{IBM Research, United States \and
  \url{https://researcher.watson.ibm.com/researcher/view.php?person=us-shinnar}}{shinnar@us.ibm.com}{}{}

\author{Vasily Pestun}{IBM Research, United States \and IH\'{E}S,
  France \and
  \url{https://pestun.ihes.fr/}}{pestun@ihes.fr}{https://orcid.org/
  0000-0001-8014-4685}{}

\authorrunning{K.\ Vajjha and B.\ Trager and A.\ Shinnar and V.\
  Pestun}

\Copyright{Koundinya Vajjha and Barry Trager and Avraham Shinnar and
  Vasily Pestun}



\relatedversion{}

\supplement{\url{https://github.com/IBM/FormalML/releases/tag/ITP2022}}

\nolinenumbers 

\EventEditors{June Andronick and Leonardo de Moura}
\EventNoEds{2}
\EventLongTitle{13th International Conference on Interactive Theorem Proving (ITP 2022)}
\EventShortTitle{ITP 2022}
\EventAcronym{ITP}
\EventYear{2022}
\EventDate{August 7--10, 2022}
\EventLocation{Haifa, Israel}
\EventLogo{}
\SeriesVolume{237}
\ArticleNo{21}

\newcommand{\R}{\mathbb{R}}

\begin{document}

\maketitle

\begin{abstract}
  Stochastic approximation algorithms are iterative procedures which
  are used to approximate a target value in an environment where the
  target is unknown and direct observations are corrupted by noise.
  These algorithms are useful, for instance, for root-finding and
  function minimization when the target function or model is not
  directly known.  Originally introduced in a 1951 paper by Robbins
  and Monro, the field of Stochastic approximation has grown
  enormously and has come to influence application domains from
  adaptive signal processing to artificial intelligence.  As an example,
  the Stochastic Gradient Descent algorithm which is ubiquitous in
  various subdomains of Machine Learning is based on stochastic
  approximation theory. In this paper, we give a formal proof (in the Coq proof assistant)
  of a general convergence theorem due to Aryeh Dvoretzky \cite{dvoretzky1956} (proven in 1956)
  which implies the
  convergence of important classical methods such as the Robbins-Monro
  and the Kiefer-Wolfowitz algorithms.  In the process, we build a
  comprehensive Coq library of measure-theoretic probability theory and
  stochastic processes.
\end{abstract}

\section{Introduction} \label{sec:introduction}
This paper presents a formal proof of Aryeh Dvoretzky's 1956 result on stochastic approximation. 

To motivate this result, let us consider a problem frequently occurring in various contexts of statistical
learning: Let $Y$ be a real-valued random variable
that depends on a parameter $x$. We may say that $P(Y|x)$ is the
probability distribution of $Y$ conditioned or dependent on a
parameter $x$.
Next, suppose we are given a function $f(y, x)$ and we
want to find $x$ that solves the equation
\begin{equation}
\label{eq:efyx}
  \mathbb{E}_P f (Y, x) = 0
\end{equation}
Moreover, assume that $P(Y|x)$ is not available to us explicitly, but
only in an implicit or sampling form, that is, we are provided a
sampling oracle which takes a parameter $x$ and returns a
sample of $Y$ drawn from $x$-dependent probability distribution
$P(Y|x)$.

\begin{example}[Kolmogorov's Strong Law of Large Numbers]
  Let $f(y, x) = y - x$,  and let $Y$ be independent of $x$. To
  solve equation (\ref{eq:efyx})  in this situation means to solve
    $\mathbb{E}[Y] = x$, that is to find the expected value $x$ of a random variable $Y$
  given an oracle from which we can sample $Y$, in other words to
  construct a statistical estimator of $\mathbb{E}[Y]$ given a series
  of samples $y_0, y_1, \dots$. The following iterative algorithm does
  the job
  \begin{equation}
    \label{eq:expectation}
    x_{n + 1} := x_{n} + a_n (y_n - x_n)
  \end{equation}
  for $n = 0, 1, 2, \dots$ where $x_{0} := 0$ and
  $a_n = \frac{1}{ n + 1}$. Indeed, the iterations
  (\ref{eq:expectation}) are equivalent to the standard sample mean
  estimator $x_n = \frac{1}{n} \sum_{k=0}^{n-1} {y_k}$.  Notice that
  the iterations (\ref{eq:expectation}) have the form
  $x_{n+1} = x_{n} + a_n f(y_n, x_n)$. The theorem that the estimator
  $x_n$ converges almost surely (with probability one) to the true
  expectation value is famously known as the ``Kolmogorov's Strong Law of Large
  Numbers'' (SLLN).
\end{example}

\begin{example}[Banach's fixed point and optimal control] Now consider the opposite example, where 
  $Y$ that depends on $x$ in a deterministic way, say $Y = g(x)$ 
  where $g: \mathbb{R} \to \mathbb{R}$
  is a certain function (and event space is a single point). In this case, when we pass $x$ to the oracle,
  the oracle deterministically returns to us the value of the function
  $g$ evaluated at $x$. In this case, solving the equation (\ref{eq:efyx}) for
  the function $f(y, x) = y - x = g(x) - x$ means solving the equation
  \begin{equation}
    g(x) = x
  \end{equation}
  If $g$ is a $\gamma$-contraction map\footnote{this means that in
    some norm $||\bullet||$ it holds that
    $||g(x) - g(x')|| < \gamma || x - x' ||$ for all $x, x'$ with
    $\gamma < 1$} the standard proof of the Banach fixed point theorem
   tells us that the iterations
  \begin{equation}
\label{eq:banach}
    x_{n+1} := x_{n} + a_n (g(x_n) - x_n)
  \end{equation}
\end{example}
for a suitable choice of $a_n$, for example
$a_n = \frac{1}{n+1}$, form a Cauchy sequence $x_1, x_2, \dots$
that converges to the fixed point of the map
$g: \mathbb{R} \to \mathbb{R}$. A variation of this process is applied
to solve Bellman's equation for optimal control of Markov Decision Process (MDP) where
$\gamma$-contraction map $g$ comes from Bellman's optimality operator
for MDPs with discount parameter $0 < \gamma < 1$.

\begin{example}[Stochastic gradient descent] Now, as a variation of
  (\ref{eq:efyx}), suppose that we want to find $x$ that minimizes the
  expectation value $\mathbb{E} [L(Y, x)]$ of a certain 
  loss function $L(Y, x)$ in a context where $Y$ is sampled by an oracle
  from an $x$-dependent probability distribution. 
  Assuming that $\mathbb{E} [L(Y, x)]$
  is a locally convex analytic function, finding a local minimum is
  equivalent to solving the stationary point equation
  \begin{equation}
    \nabla_x \mathbb{E} [L(Y, x)] = 0
  \end{equation}
  Since $\nabla_x$ is a linear operator, the above equation is equivalent to $\mathbb{E} [ \nabla_{x} L(Y, x) ] = 0$
  and therefore is again an example of the equation (\ref{eq:efyx}) with $f(y,x) := - \nabla_{x} L(y, x)$.
  The iterative sequence
  \begin{equation}
\label{eq:SGD}
    x_{n+1} := x_{n} - a_n \nabla_{x} L(y_n, x_n)
  \end{equation}
  is known as stochastic gradient descent. This algorithm is a
  typical component of most of machine learning algorithms that search
  for a parameter $x$ that minimizes the expected value of the loss
  function $L(Y, x)$ given samples of $Y$.\footnote{In the context of supervised
    learning, $y$ will stand for $(y_{in}, y_{out})$ tuples sampled
    from training data, and $x$ stands for the model parameters,
    e.g. neural network weights. If $N_{x}: y_{in} \mapsto y_{out}$ is
    a neural network, then with a quadratic supervised loss one
    normally  takes $L(y, x) := (y_{out} - N_{x}(y_{in}))^2$
    where $y=(y_{in}, y_{out})$} Under suitable conditions on
  $f(y, x) = -\nabla_{x} L(y,x)$ and the parameters $a_n$ (for
  example, $a_n = \frac{1}{n+1}$ would satisfy all required
  assumptions) one can prove convergence of (\ref{eq:SGD}) to the
  critical point of the loss function $L(Y, x)$.
\end{example}  

These three examples demonstrate the ubiquity of the problem
(\ref{eq:efyx}), and many more applications could be mentioned in a
longer report.

In all these cases the solution of the problem (\ref{eq:efyx}) has the form
\begin{equation}
\label{eq:stoch-approx}
  x_{n+1} := x_{n} + a_n f(y_n, x_n)
\end{equation}
and is called \emph{a stochastic approximation algorithm}.

A large body of literature explored different versions of assumptions
on the domain of variables, on the function $f(y, x)$ and on the
step-sizes (learning rates) $a_n$, under which the convergence of $x_{n}$ could be
proven in various senses: as convergence in $L^2$, as convergence in
probability, as convergence with probability 1.\footnote{The notion
  of ``convergence with probability 1'' is the same as the notion of
  ``convergence almost surely'', but different from the
  notion of ``convergence in probability'', which is much weaker.}

Robbins and Monro introduced in \cite{robbins1951stochastic}
the field of Stochastic Approximation by proving the $L^2$ convergence of
the process (\ref{eq:stoch-approx}) for $f(y, x) = b - y$ to the value $x$
that solves the equation  $\mathbb{E}[Y](x) = b$. Note that we write $\mathbb{E}[Y](x)$ to indicate that $x$ occurs as a parameter in the distribution of $Y$. For this theorem, Robbins and Monro assumed 
\begin{equation}
\label{eq:learning-rate}
a_n \to 0, \qquad   \sum_{n=1}^{\infty} a_n = \infty, \qquad   \sum_{n=1}^{\infty} a_n^2 < \infty,
\end{equation}
that $Y$ is bounded with probability 1, and that the function
$M(x) := \mathbb{E}[Y](x)$ is (i) non-decreasing, (ii) the solution $x_{*}$ of
$M(x) = b$ exists, and (iii) the derivative at the solution is
positive $M'(x)|_{x = x_{*}} > 0$.

Kiefer and Wolfowitz \cite{10.1214/aoms/1177729392} took a similar
approach but considered the problem of estimating the parameter $x$ where the function
$M(x)$ has a maximum, and proved convergence in probability.

Wolfowitz \cite{10.2307/2236689} weakened the assumption of
Robbins-Monro about boundedness of $Y$: instead his version assumes only
that the variance of $Y$ is bounded uniformly over $x$, and $M(x)$ is
bounded, and with those assumptions Wolfowitz proves convergence in
probability.

Blum \cite{10.1214/aoms/1177728794} weakened further the assumptions
of Robbins-Monro and Wolfowitz and proved a substantially stronger
result, namely that the iterative sequence (\ref{eq:stoch-approx}) (with $f(x_n,y_n) = b - y_n$) converges with probability
1. Blum requires the variance of $Y$ be uniformly bounded over $x$,
but he allows the expectation value $M(x) = \mathbb{E}[Y](x)$ to be
bounded by a linear function of $x$ 
\begin{equation}
\label{eq:linear-bound}
  |M(x)| \leq A |x| +  B \quad A, B \geq 0
\end{equation}
instead of a constant.  Blum's proof is based on a version of
Kolmogorov's inequality adopted in a suitable way by Lo\`eve
\cite{zbMATH03068821} where instead of series of independent random
variables, a certain dependence was allowed but constrained by a
conditional expectation value. This extension of Kolomogorov's
inequality to the conditional situation was related to earlier works
of Borel, L\'evy and Doob about convergence with probability 1 of
certain stochastic processes.

Finally, the most general form of stochastic approximation was
formulated by Dvoretzky \cite{dvoretzky1956}. In the original Robbins-Monro
stochastic approximation (\ref{eq:stoch-approx}), the next value
$x_{n+1}$ is determined through the previous value $x_{n}$ and the
sample $y_n$. Dvoretzky allowed more general estimator algorithms in
which $x_{n+1}$ is determined through a certain function that can take
as arguments complete history of all previous values
$x_{1}, \dots, x_{n}$ and the current sample $y_{n}$.

Concretely, let $T_{n}: \mathbb{R}^{n} \to \mathbb{R}$ be a
real-valued function of $n$-variables. Consider the stochastic process
\begin{equation}
\label{eq:full_history}
  x_{n+1} := T_{n}(x_1, \dots, x_n) + W_n
\end{equation}
where $W_1, W_2, \dots$ are random variables, with $W_n$ dependent on
the previous history $X_1, \dots, X_n$ such that
\begin{equation}
\label{eq:noise}
\mathbb{E}(W_n|~x_1, \dots, x_n) = 0
\end{equation}

Another way to formulate Dvoretzky's  setup is to say that for any sequence of
random variables $X_1, X_2, \dots$ where we have conditional
probability distribution of $X_{n+1}$ dependent on the complete
history $x_1, \dots, x_n$, and then \emph{define}
\begin{equation}
  \begin{aligned}
    T(x_1, \dots, x_n) & \stackrel{def}{=} \mathbb{E}[X_{n+1} | x_1, \dots, x_n] \\
    W_n & \stackrel{def}{=} X_{n+1} - \mathbb{E}[X_{n+1} | x_1, \dots, x_n]
  \end{aligned}
\end{equation}
in this way we automatically get the relation (\ref{eq:full_history})
with noise terms $W_n$ that satisfy (\ref{eq:noise}).

For example, in the Robbins-Monro version we take (\ref{eq:stoch-approx})
with $f(y,x) = b - y$ which gives
$X_{n+1} := X_{n} + a_n (b - Y_n)$ and hence the Robbins-Monro
process is a specialization of Dvoretzky's process with
\begin{equation}
  \label{eq:RM-Dvoretzky}
  \begin{aligned}
& T(x_{1}, \dots, x_{n}) := x_{n} + a_n (b - M(x_n)) \\
& W_{n} := a_n(M(x_n) - Y_n)
  \end{aligned}
\end{equation}
whereas before in the context of Robbins-Monro we had $M(x_n) :=\mathbb{E}(Y_n|x_n)$. 

To prove his result, Dvoretzky assumed that:
\begin{enumerate}
\item there exists a point $x_{*}$ such that 
\begin{equation}
\label{eq:T-bound}
  |T_n(x_1, \dots, x_n) - x_{*}| \leq \max(\alpha_n, (1 + \beta_n) | x_n - x_{*}| - \gamma_n)
\end{equation}
where $\alpha_n, \beta_n, \gamma_n$ are sequences of non-negative real numbers with
\begin{align}
\label{eq:alpha-beta-gamma}
  \alpha_n \to 0 \\
  \sum_n \beta_n < \infty \\
  \sum_n \gamma_n = \infty 
\end{align}
\item The cumulative variance of the noise terms $W_n$ is bounded
  \begin{equation}
\label{eq:comul-noise}
    \sum_{n=1}^{\infty} \mathbb{E} [W_n^2] < \infty, \qquad \mathbb{E} [W_n | X_1, \dots, X_n] = 0
  \end{equation}
\end{enumerate}
and proved that the iterative sequence (\ref{eq:full_history})
converges with probability 1 to the fixed point $x_{*}$.

The Robbins-Monro theorem in its strongest form (that is, under the
weakest assumptions of Blum (\ref{eq:linear-bound})) becomes an easy
consequence of Dvoretzky theorem.  We only have to check that given the
assumptions of Blum we can apply Dvoretzky. First, Blum's assumption
that the variance of $Y_n$ is bounded by, say, $\sigma^2$ for all $x$
and $n$, given the relation (\ref{eq:RM-Dvoretzky}), implies
$\sum_{n=1}^{\infty} \mathbb{E}[W_n^2] = \sum_{n=1}^{\infty} a_n^2
\sigma^2 < \infty$, and therefore Dvoretzky's assumption
(\ref{eq:comul-noise}) about limited cumulative variance of his noise
terms holds. Second, given Blum's $M(x)$ in equation (\ref{eq:linear-bound}), we will
construct $\alpha_n, \beta_n, \gamma_n$ that satisfy
(\ref{eq:alpha-beta-gamma}) and such that the bound on the operator $T$ in 
(\ref{eq:T-bound}) holds. To do that, first choose a real-valued
series $\{\rho_n\}$ with $\rho_n > 0$ and $\rho_n \to 0$ such
that\footnote{for example, if we start with $a_n = \frac 1 {n + 1}$
  take $\rho_n = \frac{1}{\log(n + 1)}$, in general take
  $\rho_n^{-1} = \sum_{k=1}^{n} a_n$, (see \cite{abel1828note}).
} $\sum_{n} \rho_n a_n = \infty$.
For simplicity assume, by a change of coordinates, that the
fixed point $x_{*} = 0$. Assuming that $M(x)$ is regular and monotonic
in a neighborhood of $x_{*}$ there is an inverse map $M^{-1}$ and then
we define the sequence $\{\eta_{n}\} = \{M^{-1}(\rho_n)\}$ for
sufficiently small $\rho_n$. Next, define (for sufficiently large $n$)
\begin{equation}
  \begin{aligned}
    &\alpha_n := \max(\eta_n, B a_n) \\
    &\beta_n := 0 \\
    &\gamma_n := a_n \rho_n
  \end{aligned}
\end{equation}
A case-by-case check for $|x| \leq \eta_n$ and for $|x| > \eta_n$
shows that Dvoretzky's bound on (\ref{eq:T-bound}) holds given the
relation (\ref{eq:RM-Dvoretzky}).

One universal theme passing through the various versions of stochastic
approximation convergence theorems is the choice of the scheduling of
the step-sizes (or learning rates) $a_n$.

In the Robbins-Monro scheduling assumption (\ref{eq:learning-rate}), it is
clear that the step-sizes have to converge to zero (otherwise the
model would fluctuate and never converge to the exact solution). The second assumption
$\sum_{n=1}^{\infty} a_n = \infty$ that says that the rates
should not converge to zero too fast is also sensible, as otherwise it
is easy to imagine a learning schedule with $a_n$ dropping to
zero so fast that the iterative process does not reach the fixed point $x_{*}$ from an initial point
$x_0$ (for a concrete example, see \cite[pp. 5]{chen2006stochastic}). The third assumption, $\sum_{n=1}^{\infty} a_n^2 < \infty$, is
more subtle and technical, it primarily ensures that even in
situations when the noise-terms have self-correlation they would not
move the iterative process out of its track of converging with
probability 1 to the exact fixed point.  In certain situations, a
slower decreasing of learning rate schedule still leads to
convergence, and is faster in practice.

As the above discussion has shown, the Robbins-Monro paper spawned 
a huge literature on the analysis and applications of such stochastic algorithms. 
This is because the problem of estimating unknown parameters of a model from 
observed data is quite a fundamental one, with variants of this problem appearing in
one form or another in control theory, learning theory and other
fields of engineering.  

Because of the pervasive reach of stochastic approximation methods,
any serious formalization effort of an algorithm involving parameter 
estimation when the underlying model is unknown will
eventually have to contend with formalizing tricky stochastic
convergence proofs.  We chose to formalize Dvoretzky's theorem as it
implies the convergence of both the Robbins-Monro and Kiefer-Wolfowitz
algorithms, various stochastic gradient descent agorithms and various
reinforcement learning algorithms such as Q-learning based on
Bellman's optimality operator.

\begin{remark*}
  Throughout the text which follows, hyperlinks to theorems, definitions
  and lemmas which have formal equivalents in the Coq development are
  indicated by a $\coqtop$. Our formalization
  is open-source and is
  available at~\url{https://github.com/IBM/FormalML}.
\end{remark*}

\section{Dvoretzky's Theorem}
After Dvoretzky's original publication \cite{dvoretzky1956} of his
theorem and several very useful extensions, several shorter proofs
have been proposed.  A simplified proof was published by Wolfowitz
\cite{10.1214/aoms/1177728082} who like Blum relied on the conditional
version of Kolmogorov's law exposed by Lo\`eve
\cite{zbMATH03068821}. A third, more simplified proof was published by
Derman and Sacks \cite{derman1959dvoretzky}, who again relied on the
conditional version of Kolomogorov's law, streamlined the chain of
inequality manipulations with Dvoretzky's bounding series parameters
$(\alpha_n, \beta_n, \gamma_n)$ and used Chebyshev's inequality and the
Borel-Cantelli lemma to arrive at a very short proof. Robbins and
Siegmund generalized the theorem to the context where the variables
take value in generic Hilbert spaces using the methods of supermartingale
theory \cite{ROBBINS1971233}, as did Venter \cite{10.1214/aoms/1177699145}. For
a survey see Lai \cite{10.2307/3448398}.  Dvoretzky himself published a revisited
version in \cite{DVORETZKY1986220}.

We have chosen to formalise the proof following Derman and Sacks
\cite{derman1959dvoretzky} as this version appeared to us as being the
shortest and most suitable to formalize using constructions from
our library of formalized probability theory.

In this paper we present complete formalization of the scalar version
of Dvoretzky's theorem, with random variables taking value in $\R$.

Here is a full statement of Dvoretzky's theorem: 

\begin{theorem}[Regular Dvoretzky's Theorem~\coqdef{QLearn.Dvoretzky}{Dvoretzky_DS_simple_vec_theta}\label{th:Dvoretzky}] 
  Assuming the following:
  \begin{description}
 \item[$\mathsf{H}_1:$] Let $(\Omega, \mathcal{F}, P)$ be a probability space

 \item[$\mathsf{H}_2:$] For $n = 1,2, \dots$
  
 \item[$\mathsf{H}_3:$] Let $\mathcal{F}_{n}$ be an increasing sequence of sub $\sigma$-fields
  of $\mathcal{F}$

 \item[$\mathsf{H}_4:$] Let $X_{n}$ be $\mathcal{F}_n$-measurable random
  variables taking values in $\R$.

 \item[$\mathsf{H}_5:$] Let $T_{n}: \R^{n} \to \R$
  be a measurable function

 \item[$\mathsf{H}_6:$] Let $W_{n}$ be $\mathcal{F}_{n+1}$-measurable random variables
  taking values in $\R$ such that
      \[\quad X_{n+1} = T(x_1, \dots, x_n) + W_{n}\]
 \item[$\mathsf{H}_7:$] $ \mathbb{E}(W_n | \mathcal{F}_n) = 0$

 \item[$\mathsf{H}_8:$]  $  \sum_{n=1}^{\infty} \mathbb{E} W_n^2 < \infty$

 \item[$\mathsf{H}_9:$] Let $\alpha_n, \beta_n, \gamma_n$ be a series of real numbers such
  that

 \item[$\mathsf{H}_{10}:$]  $\alpha_n \geq 0$

 \item[$\mathsf{H}_{11}:$] $\beta_n \geq 0$

 \item[$\mathsf{H}_{12}:$] $\gamma_n \geq 0$

 \item[$\mathsf{H}_{13}:$] $\lim_{n = \infty} \alpha_n = 0$

 \item[$\mathsf{H}_{14}:$] $\lim_{n = \infty} \sum_{k=1}^{n} \beta_k < \infty$

 \item[$\mathsf{H}_{15}:$] $\lim_{n = \infty} \sum_{k=1}^{n} \gamma_k = \infty$

 \item[$\mathsf{H}_{16}:$] Let $x_{*}$ be a point in $\R$ such that for all $n = 1,2,\dots$
  and for all $x_1, \dots, x_n \in \R$,
  \[| T_n(x_1, \dots, x_n) - x_{*} | \leq \max(\alpha_n, (1 + \beta_n)|x_n - x_{*}| - \gamma_n)\]
\end{description}
Then the sequence of random variables $X_1, X_2, \dots $ converges with probability 1
  to $x_{*}$: 
 \[ \quad \quad  P\{\lim_{n=\infty} X_{n} = x_{*}\} = 1 \]
\end{theorem}

An increasing sequence $\mathcal{F}_n$ of sub-$\sigma$-fields of $\mathcal{F}$ (a filtration) formalizes a notion of
a discrete stochastic process moving forward in time steps $n$, where
$\mathcal{F}_n$ formalizes the history of the process up to the time
step $n$. Assuming an $\mathcal{F}_n$-measurable random variable $X_n$
means assuming a stochastic variable $X_n$ that is included into the history
up to the time step $n$.

  We have also formalized the extended version of Dvoretzky's theorem
  in which $\alpha_n, \beta_n, \gamma_n$ are promoted to real valued functions 
  and $T_n$ is promoted to be an $\mathcal{F}_n$-measurable random variable.
  The hypotheses that have been  modified  in the extended version are marked by the symbol $\star$ below:

  \begin{theorem}[Extended Dvoretzky's theorem~\coqdef{QLearn.Dvoretzky}{Dvoretzky_DS_extended_simple_vec_theta}\label{thm:extended-dvor}] Assuming the following:
  \begin{description}
  \item [$\mathsf{H}_1:$] Let $(\Omega, \mathcal{F}, P)$ be a probability space

  \item[$\mathsf{H}_2:$] For $n = 1,2, \dots$
  
  \item[$\mathsf{H}_3:$] Let $\mathcal{F}_{n}$ be an increasing sequence of sub $\sigma$-fields
  of $\mathcal{F}$

  \item[$\mathsf{H}_4:$] Let $X_{n}$ be $\mathcal{F}_n$-measurable random
  variables taking values in $\R$.

  \item[$\star \mathsf{H}_5:$] Let $T_{n}$ be $\mathcal{F}_{n}$-measurable $\R$-valued random variable

  \item[$\mathsf{H}_6:$] Let $W_{n}$ be $\mathcal{F}_{n+1}$-measurable $\R$-valued random variables
   such that:  \[ X_{n+1} = T(x_1, \dots, x_n) + W_{n}\]

  \item[$\mathsf{H}_7:$] $\mathbb{E}(W_n | \mathcal{F}_n) = 0$

  \item[$\mathsf{H}_8:$]  $ \sum_{n=1}^{\infty} \mathbb{E} W_n^2 < \infty$

  \item[$\star \mathsf{H}_9:$] Let $\alpha_n, \beta_n, \gamma_n : \Omega \to \R$ be functions\footnote{Technically, Dvoretzky in his revisited paper \cite{DVORETZKY1986220} requires  $\alpha_n, \beta_n, \gamma_n$ to be $\mathcal{F}_{n}$-measurable, but this assumption wasn't actually used in the proof, so we have omitted it.} such that:

  \item[$\mathsf{H}_{10}:$] $\alpha_n \geq 0$

  \item[$\mathsf{H}_{11}:$] $\beta_n \geq 0$

  \item[$\mathsf{H}_{12}:$] $\gamma_n \geq 0$

  \item[$\star \mathsf{H}_{13}:$] $\lim_{n\to\infty} \alpha_n  = 0$ with probability 1

  \item[$\star \mathsf{H}_{14}:$] $\lim_{n \to \infty} \sum_{k=1}^{n} \beta_k < \infty$ with probability 1 

  \item[$\star \mathsf{H}_{15}:$] $\lim_{n \to \infty} \sum_{k=1}^{n} \gamma_k = \infty$ with probability 1

  \item[$\mathsf{H}_{16}$:] Let $x_{*}$ be a point in $\R$ such that for all $n = 1,2,\dots$
  and for all $x_1, \dots, x_n \in \R$ we have:
      \begin{equation}\label{eq:Tn-bnd}
  | T_n(x_1, \dots, x_n) - x_{*} | \leq \max(\alpha_n, (1 + \beta_n)|x_n - x_{*}| - \gamma_n)
    \end{equation}
\end{description}
Then the sequence of random variables $X_1, X_2, \dots $ converges with probability 1
to $x_{*}$: 
  \begin{equation*}
    \quad \quad P\{\lim_{n\to\infty} X_{n} = x_{*}\} = 1
  \end{equation*}
\end{theorem}

We now turn to describing the formalization of the above theorems. 
First, we give a description of our comprehensive supporting Probability Theory library in \cref{sec:form-prob} (which may be of independent interest),
then we shall give an overview of the proof of \cref{th:Dvoretzky} in \cref{sec:proof-overview}, and 
finally detail the variants of this 
theorem we have formalized in \cref{sec:variants}.

\input{sec3_2_prob_overview.tex}
\input{sec3_1_form_overview.tex}

\section{Related work}
\input{sec4_related_work.tex}

\section{Applications \& Future Work}

\input{sec5_app_future.tex}




\bibliography{Bibliography2.bib}

\appendix

\end{document}

%% file: sec3_2_prob_overview.tex

\section{Formalized Probability Library}\label{sec:form-prob}



Our formalization of both Dvoretzky theorems is built on top of our
general library of formalized Probability Theory. In particular, we
are not restricted to discrete probability measures.

\subsection{$\sigma$-Algebras and Probability Spaces}\label{sec:form-prob-event}

We first introduce
\coqe{pre_event}s~\coqdef{ProbTheory.Event}{pre_event} which are just
subsets of a type \coqe{T} i.e., maps \coqe{T -> Prop}. Then we define
$\sigma$-algebras, \coqe{SigmaAlgebra(T)}~\coqdef{ProbTheory.Event}{SigmaAlgebra}, as
collections of \coqe{pre_event}s which are closed under countable
union and complement and include the full subset of all elements in
\coqe{T}:
\begin{coq}
Class SigmaAlgebra (T : Type) :=
    {
    sa_sigma : pre_event T -> Prop;
    sa_countable_union (collection: nat -> pre_event T) :
      (forall n, sa_sigma (collection n)) ->
      sa_sigma (pre_union_of_collection collection);
    sa_complement (A:pre_event T) :
      sa_sigma A -> sa_sigma (pre_event_complement A) ;
    sa_all : sa_sigma pre_Ω
    }.
\end{coq}

Then, we label \coqe{pre_event}s which are members of a $\sigma$-algebra as
\coqe{event}s~\coqdef{ProbTheory.Event}{event}. Special $\sigma$-algebras,
like that generated by a set of
\coqe{pre_event}s~\coqdef{ProbTheory.SigmaAlgebras}{generated_sa} and the
Borel $\sigma$-algebra~\coqdef{ProbTheory.BorelSigmaAlgebra}{borel_sa},
are constructed as usual.

One interesting feature of the formalization of both of these
is that they are both provided with alternative characterizations,
which is useful for using the definitions.  For the borel $\sigma$-algebra, we define two variants:
\coqe{borel_sa}~\coqdef{ProbTheory.BorelSigmaAlgebra}{borel_sa},
defined as the $\sigma$-algebra generated by the half-open intervals,
and
\coqe{open_borel_sa}~\coqdef{ProbTheory.BorelSigmaAlgebra}{open_borel_sa},
defined as the $\sigma$-algebra generated by the open sets.  After
proving that the definitions yield the same
$\sigma$-algebra~\coqdef{ProbTheory.BorelSigmaAlgebra}{sa_borel_open_le_equiv},
we can choose which definition is simpler to work with in a given
context, simplifying some proofs.

For the definition of $\sigma(X)$, the $\sigma$-algebra generated by a
set $X$, we start with the standard
definition~\coqdef{ProbTheory.SigmaAlgebras}{generated_sa}: the
intersection~\coqdef{ProbTheory.SigmaAlgebras}{sigma_algebra_intersection}
of the set of $\sigma$-algebras that contain
$X$~\coqdef{ProbTheory.SigmaAlgebras}{all_included}.  This is useful,
but as it is non-constructive, it lacks a convenient induction
principle.  As an alternative, we define the explicit closure of a set
of
\coqe{event}s~\coqdef{ProbTheory.SigmaAlgebras}{prob_space_closure},
built by starting with the set (augmented by $\Omega$), and repeatedly
adding in complements and countable unions.  In Coq, this is naturally
defined using an inductive data type.  This closure is then shown to
be (a $\sigma$-algebra~\coqdef{ProbTheory.SigmaAlgebras}{closure_sigma_algebra} and)
equivalent to
$\sigma(X)$~\coqdef{ProbTheory.SigmaAlgebras}{generated_sa_closure}.

While definitions generally use the standard definition of
$\sigma(X)$, some theorems are more easily proven by switching to the
equivalent closure-based characterization.  This enables induction,
providing an easy way to extend a property on the generating set to
the generated $\sigma$-algebra, by showing that complements and
countable unions preserve the property in question.

Next, we introduce probability spaces
\coqdef{ProbTheory.ProbSpace}{ProbSpace} over a $\sigma$-algebra,
equipped with a measure mapping each \coqe{event} to a real number $r$, such
that $0 \le r \le 1$.
\begin{coq}
Class ProbSpace {T : Type} (σ : SigmaAlgebra T) :=
{
  ps_P : event σ -> R;
  ps_proper :> Proper (event_equiv ==> eq) ps_P ;
  ps_countable_disjoint_union (collection: nat -> event σ) :
  (* Assume: collection is a subset of Sigma and
     its elements are pairwise disjoint. *)
    collection_is_pairwise_disjoint collection ->
    sum_of_probs_equals ps_P collection (ps_P (union_of_collection collection));
  ps_one : ps_P Ω = R1;
  ps_pos (A:event σ): (0 <= ps_P A)
}.
\end{coq}

The usual properties of probability spaces, such as
monotonicity~\coqdef{ProbTheory.ProbSpace}{ps_sub},
complements~\coqdef{ProbTheory.ProbSpace}{ps_complement}, and
non-disjoint unions~\coqdef{ProbTheory.ProbSpace}{ps_union}, are
verified.

\subsection{Almost Everywhere}\label{sec:form-prob-almost}

Having defined probability spaces, we can introduce a commonly used
assertion in probabilistic proofs: that a certain property holds
\emph{almost everywhere} on a probability space. By this we mean the set of
points where the property holds includes a measurable event of measure
1.  We define a predicate
\coqe{almost}~\coqdef{ProbTheory.Almost}{almost} to indicate
propositions which hold almost everywhere.  It is parameterized by a
probability space and proposition on that space.

\begin{coq}
  Definition almost {Ts:Type} {dom: SigmaAlgebra Ts}(prts: ProbSpace dom) (P:Ts -> Prop)
    := exists E, ps_P E = 1 /\ forall x, E x -> P x.
\end{coq}

We have introduced machinery to make it more convenient to reason
about \coqe{almost} propositions. For example, if we want to show that
\coqe{almost P -> almost Q -> almost R}, we reduce the proof to
showing that
\coqe{almost (P -> Q -> R)}~\coqdef{ProbTheory.Almost}{almost_impl},
which itself is implied by
\coqe{P -> Q -> R}~\coqdef{ProbTheory.Almost}{all_almost}. Usual
theorem proving tools can then be used.

On top of the basic \coqe{almost} definition, we defined
\coqe{almostR2}~\coqdef{ProbTheory.Almost}{almostR2}, which says that
a binary relation holds \coqe{almost} everywhere.
\begin{coq}
 Definition almostR2 (R:Td->Td->Prop) (r1 r2:Ts -> Td) : Prop
   := almost (fun x => R (r1 x) (r2 x)).
 \end{coq}

 This is useful, since it inherits many properties from the base
 relation (e.g. it is a preorder if the base relation
 is~\coqdef{ProbTheory.Almost}{almostR2_pre}), and simplifies
 definitions.

\subsection{Measurability and Expectation}\label{sec:form-prob-expect}

We next introduce the concept of measurable functions with respect to
two $\sigma$-algebras.  Since we are focusing on probability spaces,
we call these measurable functions
\coqe{RandomVariable}s~\coqdef{ProbTheory.RandomVariable}{RandomVariable}.
\begin{coq}
(* A random variable is a mapping from a probability space to a sigma algebra. *)
Class RandomVariable {Ts:Type} {Td:Type}
      (dom: SigmaAlgebra Ts)
      (cod: SigmaAlgebra Td)
      (rv_X: Ts -> Td)
  := (* for every element B in the sigma algebra, the preimage
     of rv_X on B is an event in the probability space *)
    rv_preimage_sa: forall (B: event cod), sa_sigma (event_preimage rv_X B).
\end{coq}

In order to define the \coqe{Expectation} of a \coqe{RandomVariable},
we follow the usual technique of first treating the case of finite
range
functions~\coqdef{ProbTheory.SimpleExpectation}{SimpleExpectation},
then extending to nonnegative
functions~\coqdef{ProbTheory.Expectation}{NonnegExpectation}
(resulting in an extended real) and then to general random
variables. In the general case, the expectation is the difference of
the expectation of the positive and negative parts of a random
variable.~\coqdef{ProbTheory.Expectation}{Expectation} Exceptions are
handled using the Coq \coqe{option} type. For example, the difference
of the expectations of the positive and negative parts of a random
variable is not defined if they are both the same infinity. This
exception is captured by allowing the difference to be \coqe{None} in
that case.  A well defined Expectation yields \coqe{Some r}, for some
value in Coquelicot's \coqe{Rbar} type~\cite{Coquelicot}.  This represents a value in
the extended reals: either a \coqe{Finite} real value, or positive or
negative infinity (\coqe{p_infty} or \coqe{m_infty}).

\begin{coq}
  Definition Expectation (rv_X : Ts -> R) : option Rbar :=
    Rbar_minus' (NonnegExpectation (pos_fun_part rv_X))
                (NonnegExpectation (neg_fun_part rv_X)).
\end{coq}

Originally our results about \coqe{Expectation} were for random
variables taking images in the reals, but as we introduced limiting
processes we needed to extend our definition to random variables taking
values in the extended reals (\coqe{Rbar}).

This requires extending the support for limits in Coquelicot, allowing
for sequences of functions over the extended
reals~\coqdef{utils.ELim_Seq}{is_Elim_seq}.  The approach we took was
to copy over all the definitions and lemmas in Coquelicot's
\coqe{Lim_seq} module, extending them as appropriate, and re-proving
them.  A few changes were made, such as defining the extended version
of \coqe{is_lim_seq}~\coqdef{utils.ELim_Seq}{is_Elim_seq} to hold when
the inf and sup sequence limits coincide.  The original definition
uses filters, and is problematic to extend to the extended reals,
since they do not form a uniform space.  Pleasantly, however, almost
all of the lemmas continue to hold with minor modification.

The above construction of Expectation and its properties (including
linearity~\coqdef{ProbTheory.Expectation}{Expectation_scale}\coqdef{ProbTheory.Expectation}{Expectation_sum},
the monotone convergence
theorem~\coqdef{ProbTheory.Expectation}{monotone_convergence}, and
other standard results) are then generalized and proven for this
generalization to functions whose image is the extended
reals~\coqdef{ProbTheory.RbarExpectation}{Rbar_Expectation}.

On top of our general definition of \coqe{Expectation}, we define the
\coqe{IsFiniteExpectation} property, which asserts that a function has
a well-defined, finite
expectation~\coqdef{ProbTheory.RandomVariableFinite}{IsFiniteExpectation}\coqdef{ProbTheory.RbarExpectation}{Rbar_IsFiniteExpectation}.
For functions that satisfy this property, we can define their
\coqe{FiniteExpectation}~\coqdef{ProbTheory.RandomVariableFinite}{FiniteExpectation}\coqdef{ProbTheory.RbarExpectation}{Rbar_FiniteExpectation},
which returns their (real) expectation.  This simplifies working with
such functions, and avoids otherwise necessary side-conditions on
properties such as linearity~\coqdef{ProbTheory.RandomVariableFinite}{FiniteExpectation_scale}\coqdef{ProbTheory.RandomVariableFinite}{FiniteExpectation_plus}.

\subsection{$L^p$ Spaces}\label{sec:form-prob-lp}

Using these building blocks, we can define $L^p$ spaces, which are the
space of measurable functions where the $p$-th power of its absolute
value has finite
expectation~\coqdef{ProbTheory.RandomVariableLpR}{IsLp}.
\begin{coq}
Definition IsLp {Ts} {dom: SigmaAlgebra Ts} (prts: ProbSpace dom) (n:R) (rv_X:Ts->R)
    := IsFiniteExpectation prts (rvpower (rvabs rv_X) (const n)).  
\end{coq}

This space is then quotiented, identifying functions that are equal
\coqe{almost} everywhere (see
\cref{sec:form-prob-almost})~\coqdef{ProbTheory.RandomVariableLpR}{LpRRVq}.
We use a quotient construction~\coqdef{utils.quotient_space}{quot}
that avoids needing axioms beyond those already proposed in Coq's
standard libraries\footnote{Specifically, we use functional and
  propositional extensionality as well as constructive definite
  description (also known as the axiom of unique choice).}.  This
quotienting operation is required in order to define a norm on the
space (defined as the $p$-th root of the \coqe{Expectation} of the
absolute value of the $p$-th power of the function), as having a zero
\coqe{Expectation} only implies that a non-negative function is zero
\coqe{almost} everywhere.

For nonnegative $p$, $L^p$ is shown to be a module
space~\coqdef{ProbTheory.RandomVariableLpR}{LpRRVq_ModuleSpace}.  For
$1\le p \le \infty$, it is shown to be a Banach space (complete normed
module
space)~\coqdef{ProbTheory.RandomVariableLpR}{LpRRVq_CompleteNormedModule}
\coqdef{ProbTheory.RandomVariableLinf}{LiRRVq_CompleteNormedModule}.

Furthermore, the important special case of $L^2$ is proven to be a
Hilbert space~\coqdef{ProbTheory.RandomVariableL2}{L2RRVq_Hilbert},
where the inner product of $x$ and $y$ is defined as the
\coqe{Expectation} of the product of $x$ and $y$.

\subsection{Conditional Expectation}\label{sec:form-prob-condexp}

Building on top of this work, we turn to the definition of conditional
expectation, defining it with respect to a general $\sigma$-algebra
\coqe{dom2} (the ambient $\sigma$-algebra being \coqe{dom}).  We first
postulate a relational
definition~\coqdef{ProbTheory.ConditionalExpectation}{is_conditional_expectation},
characterized by the universal property of conditional expectations:
for any event $P$ that is in the sub $\sigma$-algebra \coqe{dom2}, if
we multiply the original function and its conditional expectation by
that event's associated indicator function, we get equal expectations.

\begin{coq}
  Definition is_conditional_expectation {Ts:Type} {dom: SigmaAlgebra Ts}
      (prts: ProbSpace dom) (dom2 : SigmaAlgebra Ts)
      (f : Ts -> R) (ce : Ts -> Rbar)           
      {rvf : RandomVariable dom borel_sa f}
      {rvce : RandomVariable dom2 Rbar_borel_sa ce}
    := forall P (dec:dec_pre_event P),
      sa_sigma (SigmaAlgebra := dom2) P ->
      Expectation (rvmult f (EventIndicator dec)) =
      Rbar_Expectation (Rbar_rvmult ce (EventIndicator dec)).  
\end{coq}

Using this definition, we can show uniqueness (where equality is
almost
everywhere)~\coqdef{ProbTheory.ConditionalExpectation}{is_conditional_expectation_unique},
and many standard properties of conditional expectation, such as
linearity~\coqdef{ProbTheory.ConditionalExpectation}{is_conditional_expectation_scale}\coqdef{ProbTheory.ConditionalExpectation}{is_conditional_expectation_plus},
preservation of
\coqe{Expectation}~\coqdef{ProbTheory.ConditionalExpectation}{is_conditional_expectation_Expectation},
(\coqe{almost})
monotonicity~\coqdef{ProbTheory.ConditionalExpectation}{is_conditional_expectation_ale},
and the tower
law~\coqdef{ProbTheory.ConditionalExpectation}{is_conditional_expectation_tower}.
We also show the ``factor out''
property~\coqdef{ProbTheory.ConditionalExpectation}{is_conditional_expectation_factor_out}\label{factor_out},
which enables factoring out of a conditional expectation a random variable
that is measurable with respect to the sub $\sigma$-algebra.  In
addition, we verify its interactions with limits (e.g.  the
conditional version of the monotone convergence
theorem~\coqdef{ProbTheory.ConditionalExpectation}{is_conditional_expectation_monotone_convergence}),
and prove Jensen's lemma
\coqdef{ProbTheory.ConditionalExpectation}{is_condexp_Jensen},
bounding how convex functions affect the conditional expectation.

After having proven these properties for the
\coqe{is_conditional_expectation} relation, we still need to show that
the conditional expectation generally exists (at least for functions
that are non-negative or have finite expectation).

To do this, we build on our work on $L^p$ spaces
(Section~\ref{sec:form-prob-lp}), and in particular our proof that
that $L^2$ is a Hilbert space.  Given an $L^2$ function, this implies
that the subset of functions which are measurable with respect to a smaller
$\sigma$-algebra \coqe{dom2} forms a linear subspace.

The $L^2$ conditional
expectation~\coqdef{ProbTheory.ConditionalExpectation}{conditional_expectation_L2q}
of an $L^2$ random variable $X$ with respect to \coqe{dom2} is then
defined as the orthogonal
projection~\coqdef{ProbTheory.OrthoProject}{ortho_projection_hilbert}
of $X$ onto that subspace. For this construction and definitions of
Hilbert spaces we use the library from the formal development of the
Lax-Milgram theorem~\cite{BoldoElfic}.  Note that this definition is
for a function in the quotiented space (recall that $L^2$ is quotiented to
identify functions that are equal almost everywhere).

We can then define conditional expectation on the unquotiented space
by injecting the inputs into the quotiented space, using the
conditional expectation operator just defined on $L^2$ functions, and
then choosing a representative from the equivalence class of functions
it
returns~\coqdef{ProbTheory.ConditionalExpectation}{conditional_expectation_L2}.
This unquotienting gives insight into why most theorems about
conditional expectations only \coqe{almost} hold, as it is
defined on equivalence classes of \coqe{almost} equal functions.

Next, we extend our notion of conditional expectation to nonnegative
functions whose usual expectation is finite using the property that
$L^2$ functions are dense in $L^1$. In particular, given a nonnegative
$L^1$ function $f$, we can define an $L^2$ sequence
$g_n = \min(f, n)$. The conditional expectation of $f$ is defined as
the limit of the conditional expectation of the
$g_n$~\coqdef{ProbTheory.ConditionalExpectation}{NonNegConditionalExpectation}.

Using this definition directly has some disadvantages: it forces
essentially all theorems, including simple ones such as the result
being non-negative, or that the conditional expectation is the
identity operation on functions that are already measurable with
respect to the sub $\sigma$-algebra \coqe{dom2}, be only \coqe{almost}
valid.  To address this, we wrap this definition in a
wrapper~\coqdef{ProbTheory.ConditionalExpectation}{NonNegCondexp} that
takes the function returned by the original (limit based) definition
and tweaks it slightly, producing a ``fixed'' function \coqe{almost}
equivalent to the original, but where such simple properties hold
unconditionally.

Finally, we extend this to all measurable functions by taking the
difference of the nonnegative conditional expectation of its positive
and negative
parts~\coqdef{ProbTheory.ConditionalExpectation}{ConditionalExpectation}.
While this function is defined for all measurable functions, it can
only be shown to be a conditional expectation (the
\coqe{is_conditional_expectation} relation defined above) for
functions that are either
non-negative~\coqdef{ProbTheory.ConditionalExpectation}{Condexp_cond_exp_nneg}
or have finite
expectation~\coqdef{ProbTheory.ConditionalExpectation}{Condexp_cond_exp}.
Using this property, we now lift all of the properties proven above
for the relational version to our explicitly defined version, verifying
that it satisfies all the expected properties.  For convenience, we
also provide a wrapper definition
\coqe{FiniteConditionalExpectation}~\coqdef{ProbTheory.ConditionalExpectation}{FiniteConditionalExpectation},
which assumes that the function has finite expectation, and returns a
function whose image is in $\R$ (insted of the extended reals), and
lift all the expected properties to it.

Connecting back to $L^p$ spaces, we can use Jensen's lemma about
convex functions to show that if a function is in $L^p$ then its
conditional expectation is as
well~\coqdef{ProbTheory.ConditionalExpectation}{FiniteCondexp_Lp},
allowing us to view conditional expectation as a
(contractive~\coqdef{ProbTheory.ConditionalExpectation}{FiniteCondexp_Lp_contractive})
operation on $L^p$ spaces.  Furthermore, we show that it minimizes the $L^2$-loss for an $L^2$ function
~\coqdef{ProbTheory.ConditionalExpectation}{FiniteCondexp_L2_min_dist}.

We chose this approach to defining conditional expectation (via an
orthonormal projection on $L^2$) since we could rely on an existing
library of Hilbert space theory \cite{BoldoElfic}, thus avoiding other tedious
constructions involving Radon-Nikodym derivatives etc.

\subsection{Filtrations and Martingales}
\label{sec:form-prob-filter-mart}

We next introduce a notion of $\sigma$-algebra
filtrations~\coqdef{ProbTheory.SigmaAlgebras}{IsFiltration}, which are
an increasing sequence of $\sigma$-algebras. We say that a sequence of
random variables $X_n$ \coqe{IsAdapted}~\coqdef{ProbTheory.RandomVariable}{IsAdapted} to a filtration
$F_n$, if each random variable of the sequence is measurable with
respect the corresponding $\sigma$-algebra.

Building on these definitions and our development of conditional
expectation, we started developing the basics of martingale
theory~\coqdef{ProbTheory.Martingale}{IsMartingale}.

Additionally, the language of filtrations and adapted processes
enables us to represent the history of a stochastic process, which is
critical for stating and verifying properties of stochastic
approximation methods.

\subsection{Additional results}
\label{sec:form-prob-add}

There are many other results proven in the library; here we highlight
two that are used in the Derman-Sacks proof: Chebyshev's inequality
and the Borel-Cantelli lemma.

Chebyshev's inequality~\coqdef{ProbTheory.Expectation}{Chebyshev_ineq_div_mean0} which states
that given a random variable $X$ and a positive constant $a$, the
probability of $\| X \| \ge a$ is less that or equal to the
expectation of $ X^2 / a^2$.
\begin{coq}
  Lemma Chebyshev_ineq_div_mean0
  (X : Ts -> R) (rv : RandomVariable dom borel_sa X) (a : posreal) :
  Rbar_le (ps_P (event_ge dom (rvabs X) a)) 
          (Rbar_div_pos 
          (NonnegExpectation (rvsqr X)) 
          (mkposreal _ (rsqr_pos a))).
\end{coq}

Another is the Borel-Cantelli lemma~\coqdef{ProbTheory.RandomVariableFinite}{Borel-Cantelli} which states
that if the sum of probabilities of a sequence of events is finite,
then the probability of all but finitely many of them occuring is 0.

\begin{coq}
  Theorem Borel_Cantelli (E : nat -> event dom) :
 (forall (n:nat), sa_sigma (E n)) ->
  ex_series (fun n => ps_P (E n)) ->
  ps_P (inter_of_collection 
          (fun k => union_of_collection 
                      (fun n => E (n + k)))) = 0.
\end{coq}

In this theorem statement, \coqe{ex_series f}, defined in Coquelicot,
assert that the infinite series of partial sums
$\lim_{n\to\infty} \sum_{0\le i\le n} f(i)$ converges to a finite limit.

\subsection{Retrospective design decisions}
\label{sec:retro_design}

In this section we discuss some of the design choices we made (and
revisited), and our retrospective opinion on their impact.  This may
be of benefit to those seeking to pursue similar projects.

Initially, we modeled events as sets (now called \coqe{pre_event}s),
accompanying them with proofs that the set was in a given
$\sigma$-algebra.  This resulted in a lot of code threading and
transforming these proofs, which was particularly painful when
reasoning about lists.  Revisiting that decision, we built up
\coqe{event}s as a subset type: a dependent pair of a \coqe{pre_event}
and a proof that it belongs in a relevant
$\sigma$-algebra~\coqdef{ProbTheory.Event}{event}.  For many simple
uses, this obviates the need for reasoning explicitly about being in a
$\sigma$-algebra.  There are still cases where explicit reasoning is
required, but this change definitely simplified the code.

We support general random variables using a typeclass which specifies the sigma algebras for the domain and range along with the function~\coqdef{ProbTheory.RandomVariable}{RandomVariable}.
An initial version of our probability library developed expectation and properties of random variables whose codomain was the reals~\coqdef{ProbTheory.BorelSigmaAlgebra}{borel_sa}. However as we proved more properties, especially limiting and convergence properties, it became clear that we needed to allow infinite values, thus to allow random variables with $\bar \R$ (the extended real numbers) as codomain~\coqdef{ProbTheory.BorelSigmaAlgebra}{Rbar_borel_sa}. However the native support for limits in the Coquelicot package allows the limiting value to be infinite, but restricts to sequences taking values in $\R$. In order to be able to take limits of random variables to $\bar \R$, we developed our own limit package extending the Coquelicot definitions and lemmas to sequences taking values in $\bar \R$~\coqdef{utils.ELim_Seq}{is_Elim_seq}. In the end, some results about $\bar \R$ valued random variables become simpler since one doesn't need to make unnatural finiteness restrictions, but on the other hand, one needs to be extra careful, since $\bar \R$ is not a field as sums and products are not always defined and operations are not associative.

As in the standard development of expectation, we first defined it for functions whose range is a finite subset of $\R$ (or $\bar \R$).
We decided to represent them as a typeclass which includes a field containing a finite list of values which includes all the values in the range of the function~\coqdef{ProbTheory.RandomVariable}{FiniteRangeFunction}.
\begin{coq}
  Class FiniteRangeFunction
        (rv_X:Ts->Td)
    := {
    frf_vals : list Td ;
    frf_vals_complete : forall x, In (rv_X x) frf_vals;
      }.
\end{coq}
We decided to allow this list to have duplicates and to contain additional values not in the range. This made several definitions more convenient, for example when defining the sum of two finite range functions, the new list of values is just the sum of all pairs of values, which is guaranteed to contain all the actual values, but can contain values which are not in the image of the sum and can contain duplicated values~\coqdef{ProbTheory.RealRandomVariable}{frfplus}.

One simplification our code makes is that we deal only with
probability spaces, rather then general measures.  This was a
pragmatic decision, as it simplifies some of the proofs (since, for
example, measures must be finite).  As our intended use is probability
theory, this mostly sufficed.  In order to define
(dependent) product spaces, we did define the rudiments of measure
theory (measures, outer measures, and inner
measures)~\coqdef{ProbTheory.Measures}{measure}, but the final construction of the product is defined only for probability spaces, since the proof
crucially relies on the monotone convergence theorem, which we have
not proven for general measure spaces.  It would clearly have been
nicer to define things more generally, and we may go back and change
things in the future, however this simplification allowed us to use
our limited resources to greater effect.


%% file: sec3_1_form_overview.tex
\section{Formalization Challenges/Overview}
We will now sketch the key pieces which go into 
the formalization of the Derman-Sacks proof. 


\subsection{Overview of the proof}\label{sec:proof-overview}
The Derman-Sacks proof relies on a number of prerequisites 
in Probability Theory and Real Analysis. For example, the proof begins
by stating that we may replace the series $\sum_n \mathbb{E}W_n^2 < \infty$
by the series $\sum_n \frac{\mathbb{E}W_n^2}{\alpha_n^2} < \infty$ where $\alpha_n \to 0$.
This statement invokes a classical theorem of du Bois-Reymond \cite{bois1873neue}
which states:

\begin{theorem}\coqdef{utils.Sums}{no_worst_converge_iff}\label{thm:du-bois}
  Let $(a_n)$ be a sequence of nonnegative real numbers. The series 
  $\sum_n a_n$ converges if and only if there is another sequence 
  of positive real numbers $(b_n)$ such that $b_n \to \infty$ and 
  $\sum_n a_n b_n < \infty$. 
\end{theorem}
In other words, this theorem states that \textit{no worst convergent series 
exists} (see \cite{ash1997neither}). This elementary theorem did require 
some effort to formalize, in part because existing proofs such as 
the one in \cite{ash1997neither} require the sequence $(a_n)$ to consist 
only of positive terms, while our application (Dvoretzky's theorem) needed 
them to be non-negative. Additionally, we had to prove convergence 
of the product series without using the integral test (as used in \cite{ash1997neither}),
because it was unavailable in our library. Our final proof of \cref{thm:du-bois} involved a case analysis in 
which we case on whether the sequence $(a_n)$ was eventually positive or not \coqdef{utils.RealAdd}{eventually_pos_dec},
and we bypassed the need to use the integral test by using an exercise 
from Rudin's \textit{Principles of Mathematical Analysis} \cite{rudin1976principles}.

The main workhorse of the Derman-Sacks proof is the sequence 
$Z_n := W_n \ \mathrm{sgn} \ T_n$. First, they apply the following 
theorem\footnote{the proof of this theorem is a modification of Theorem 
6.2.1 in Ash's \textit{Probability and Measure Theory} \cite{ash2000probability}}
to the sequence of random variables $(Z_n)$: 

\begin{theorem}[Lo\`eve \cite{zbMATH03068821}~\coqdef{FormalML.slln}{Ash_6_2_1_filter}]\label{thm:loeve}
 Let $X_1,X_2,\dots$ be a sequence of random variables adapted to a filtration 
 $(\mathcal{F}_n)_{n\in \mathbb{N}}$. 
 Assume that $\mathbb{E}[X_{n+1} \ | \ \mathcal{F}_n] = 0$ almost surely for all $n$ and also 
 that $\sum_{n=1}^\infty \mathbb{E}X_n^2$ converges. Then we have that $\sum_{n=1}^\infty X_n$ converges 
 almost surely. 
\end{theorem}
to conclude that the series $\sum_n Z_n$ converges almost surely. 
To apply this theorem we need to prove that $(Z_n)$ is adapted to the filtration $\mathcal{F}$, 
which critically uses the fact that $T_n : \mathcal{H}^n \to \mathcal{H}$ is a 
measurable function. (Here we take $\mathcal{H} = \R$.) The proof of the theorem
uses  $\mathbb{E}[X_{n+1} \ | \ \mathcal{F}_n] = 0$ to show that since the sequence is adapted,
we have  $\mathbb{E}[X_i X_j] = 0$ for all $i \neq j$. This depends on the ``factor out''
property of conditional expectation~\coqdef{ProbTheory.ConditionalExpectation}{is_conditional_expectation_factor_out} (see \cref{factor_out}).

Next, it is shown that $|Z_n| \le \alpha_n$ almost surely for sufficiently
large $n$. This argument uses the Borel-Cantelli lemma \coqdef{ProbTheory.RandomVariableFinite}{Borel_Cantelli} 
and the Chebyshev inequality \coqdef{ProbTheory.Expectation}{Chebyshev_ineq_div_mean0}, 
both of which needed a significant amount of probability theory 
to be set up (see \cref{sec:form-prob-add}). Using this bound for $Z_n$ and the 
bound for $|T_n|$ in the hypothesis, an elementary argument shows that
\[ 
|X_{n+1}| \le \max(2 \alpha_n, |T_n| + Z_n) \le \max(2 \alpha_n, (1 + \beta_n)|X_n| + Z_n - \gamma_n)
\]
almost surely for sufficiently large $n$. 

Now, the conclusion $X_{n+1} \to 0$ almost surely follows by applying the following lemma:
\begin{lemma}\coqdef{QLearn.Dvoretzky}{DS_lemma1}
  Let $\{a_n\},\{b_n\},\{c_n\},\{\delta_n\}$ and $\{\xi_n\}$ be sequences of real
  numbers such that 
  \begin{enumerate}
    \item $\{a_n\},\{b_n\},\{c_n\},\{\xi_n\}$ are non-negative
    \item $\lim_{n \to \infty} a_n = 0, \ \sum_n b_n < \infty, \ \sum_n c_n = \infty, \ \sum_n \delta_n$ converges.
    \item For all $n$ larger than some $N_0$, $\xi_{n+1} \le \max(a_n, (1 + b_n)\xi_n + \delta_n - c_n)$ \label{it:lemma1_prop}
  \end{enumerate}
  then, $\lim_{n \to \infty}\xi_n = 0$.
\end{lemma}

The proof of the lemma is somewhat unusual since it involves running an iteration backwards: the property (\ref{it:lemma1_prop}) 
is applied repeatedly to derive an inequality between $\xi_{n+1}$ and $\xi_{N}$ for $n > N > N_0$ \coqdef{QLearn.Dvoretzky}{DS_1_helper}.
Besides using several properties of infinite products and list maximums, the final convergence result is an application of Abel's
descending convergence criterion \coqdef{utils.Sums}{Abel_descending_convergence} which says if the series $\sum_n b_n$ converges, and $a_n$ is a bounded descending sequence, then the series $\sum_n a_n b_n$ also converges.

We note that our formalization is firmly within the Classical 
territory for a number of reasons: first of all, the theory of 
Real numbers within the Coq standard library (which we use) uses non-computable axioms \cite{geuvers2000constructive}. 
Secondly, while constructive measure theory and constructive analysis are both
actively researched topics (see \cite{coquand2008integrals,coquand2002metric,bishop1967foundations}) 
we are unaware if our main result (Dvoretzky's theorem) is constructively valid. 
Thirdly, as we remarked above, our proof of \cref{thm:du-bois} requires a case 
split on whether a particular sequence of real numbers is eventually zero or not, for which 
we use the axiom of constructive indefinite description.

\subsection{Variants of Dvoretzky's Theorem.}\label{sec:variants}

While Dvoretzky's theorem admits generalizations in many different ways, we chose 
to focus on formalizing the ones most suited for applications. 
\begin{enumerate}
  \item As already mentioned, we prove \cref{thm:extended-dvor} which is a generalization of 
  \cref{th:Dvoretzky} in which the sequences of numbers $\alpha_n, \ \beta_n, \ \gamma_n$
  are replaced by sequences of functions on the probability space. 
  This generalization is called the \textit{extended} Dvoretzky theorem \coqdef{QLearn.Dvoretzky}{Dvoretzky_DS_extended_alt_simple_vec_theta}.
  All conditions on the sequences $\alpha_n, \ \beta_n, \ \gamma_n$ now hold pointwise,
   almost everywhere. 
  \item   To apply \cref{thm:loeve} in the proof of \cref{th:Dvoretzky} we needed to
  prove that $(Z_n)$ is adapted to the filtration $\mathcal{F}$, which needed us to make
  assumptions on the functions $T_n$. These assumptions on $T_n$ can be modified and 
  generalized as:
  \begin{enumerate}
    \item  in the regular (non-extended) case, $T_n : \R^n \to \R$ are deterministic and measurable. \coqdef{QLearn.Dvoretzky}{Dvoretzky_DS_simple_vec}
    \item in the extended case, $T_n : \R^n \times \Omega \to \R$ are stochastic and $\mathcal{F}_n$-adapted. \coqdef{QLearn.Dvoretzky}{Dvoretzky_DS_extended_simple_vec}
  \end{enumerate} 
  Since Derman-Sacks do not explicitly state either assumption, we formalized 
  Dvoretzky's theorem under both assumptions. It should be noted that Dvoretzky's
  original paper \cite{dvoretzky1956} and his revisited paper \cite{DVORETZKY1986220} 
  treat both the above cases. 
  \item  We have also formalized a corollary of the extended Dvoretzky's theorem \coqdef{QLearn.Dvoretzky}{Dvoretzky_DS_extended_alt_simple_vec_theta} which proves that the 
  theorem holds in the context where the bound on $T$ in (\ref{eq:T-bound}) is assumed as follows with all other assumptions intact:
  \[
   | T_n(x_1, \dots, x_n) - x_{*} | \leq \max(\alpha_n, (1 + \beta_n - \gamma_n)|x_n - x_{*}|)  
  \]
  While this formulation is weaker compared to the original, it is convenient to have it for several
  applications of stochastic approximation theorems.  A proof of
  this corollary used a classical analysis result of Abel \cite{abel1828note}
  on the fact that the terms in a divergent sum-series could be multiplied by
  infinitesimally small series and the sum-series would still diverge
  \coqdef{utils.RealAdd}{no_best_diverge_iff}. This was addressed in 
  Dvoretzky's paper \cite[(5.1)]{dvoretzky1956}.
\end{enumerate}



%% file: sec4_related_work.tex
While our results are general, our intended application 
was formalizing machine learning theory, on which there is a 
growing body of work \cite{DBLP:journals/corr/abs-1911-00385,tassarotti2021formal,DBLP:journals/corr/abs-2007-06776,markovInHOL,DBLP:conf/icml/SelsamLD17,DBLP:conf/aaai/Bagnall019,DBLP:journals/jar/BentkampBK19}.
Our work is a step in this direction, providing future developers
of secure machine learning systems a library of formalized 
stochastic approximation results. Keeping this in mind, we have 
formalized different versions of our main result (Dvoretzky's theorem)
to facilitate ease of use (see \cref{sec:variants}).


For the formalization itself,
we make extensive use of the Coquelicot library of Boldo et al.
\cite{Coquelicot} and the library which proved the Lax-Milgram theorem \cite{BoldoElfic} which includes
definitions and basic properties of hilbert spaces.
There have also been quite a
few formalizations of probability theory in Coq: see Polaris \cite{tassarotti2019separation}, Infotheo \cite{affeldt2012itp},
and Alea \cite{audebaud2009proofs}.
Alea is an early work and to the best of our knowledge incompatible 
with latest versions of Coq while Infotheo and Polaris either fundamentally focus on discrete probability
theory (see \cite{affeldt2020}) or do not have the results we needed to prove Dvoretzky's theorem. 

More recently there have been two projects in Coq which formalize measure theory and Lebesgue integration.
The MathComp-Analysis project has general measure theory and integration developed on top of their
library which is an alternative to Coquelicot \cite{affeldt2021coq}.
The Numerical Analysis in Coq (coq-num-analysis) project is built on top of Coquelicot and
includes support for Lebesgue integration of nonnegative functions \cite{boldo2021coq}.
Neither of these were available at the time we began to develop our probability library. Since we depend
on Coquelicot, we could have developed on top of the coq-num-analysis library if it were available earlier.
This would have given us the added benefit of supporting general measures instead of our restriction to probability measures.
Refactoring our library to build on top of one or more of these formalizations might be a possible direction for future work.

Formal proofs about convergence of random variables (the Central Limit Theorem) have been given
in Avigad et al \cite{avigad2017formally} using the Isabelle/HOL system. Parts of Martingale theory and stochastic processes have also recently made their way into the Lean math
library \cite{mathlib2020lean}. 

To the best of our knowledge, our work presents the first formal 
proof of correctness of any theorem in Stochastic Approximation. 


%% file: sec5_app_future.tex
Our own interest in stochastic approximation began with an attempt to
extend our work on convergence proofs of (model-based) Reinforcement
Learning (RL) algorithms \cite{vajjha2021certrl} to include the
model-free case.  Model-based RL algorithms converge to an
\textit{optimal policy} (a sequence of actions which an agent should
probabilistically perform so as to maximize its expected long-term
reward) by making full use of the given transition probability
structure of the agent.  The term \textit{model-free} refers to the
fact that we have no information on how the agent performs it's
transitions but can only \textit{observe} its transitions.  As we have
emphasized above, this situation is perfectly suited for stochastic
approximation techniques.  Indeed, convergence proofs of Q-Learning (a
prominent model-free RL algorithm) appeal to standard results of
stochastic approximation (see Watkins \& Dayan \cite{watkins1992q},
Jaakkola et al. \cite{jaakkola1994convergence}), Tsitsiklis
\cite{tsitsiklis1994asynchronous}. We plan to use our formalization of 
Dvoretzky's theorem to complete a convergence proof of the Q-learning algorithm.

Additionally, as part of this process, we have built up a large
library for basic results on (general) probability spaces in Coq,
including a general definition of conditional expectation.  This
library is publically
available at~\url{https://github.com/IBM/FormalML} and open source.
We invite others to use our library and collaborate with us on
extending and enhancing it.
